\documentclass[aps,prl,amsmath,amssymb,twocolumn,floatfix,10pt,showpacs,superscriptaddress]{revtex4-1}
\usepackage{graphicx}
\usepackage{hyperref}
\usepackage{color}

\newcommand{\bs}{\;\;\;\;\;}

\newcommand{\ve}{\mathbf}

\begin{document}

\title{Magnetic Correlations in Short and Narrow Graphene Armchair Nanoribbons}
\author{Michael Golor}
\affiliation{Institut f\"ur Theoretische Festk\"orperphysik, JARA-FIT and JARA-HPC, RWTH Aachen University, 52056 Aachen, Germany}
\author{Cornelie Koop}
\affiliation{Institut f\"ur Theoretische Festk\"orperphysik, JARA-FIT and JARA-HPC, RWTH Aachen University, 52056 Aachen, Germany}
\author{Thomas C. Lang}
\affiliation{Department of Physics, Boston University, Boston, MA 02215, USA}
\author{Stefan Wessel}
\affiliation{Institut f\"ur Theoretische Festk\"orperphysik, JARA-FIT and JARA-HPC, RWTH Aachen University, 52056 Aachen, Germany}
\author{Manuel J. Schmidt}
\affiliation{Institut f\"ur Theoretische Festk\"orperphysik, JARA-FIT and JARA-HPC, RWTH Aachen University, 52056 Aachen, Germany}
\date{\today}


\begin{abstract}
Electronic states at the ends of a narrow armchair nanoribbon give rise to a pair of non-locally entangled spins. We propose two experiments to probe these magnetic states, based on magnetometry and tunneling spectroscopy, in which correlation effects lead to a striking, nonlinear response to external magnetic fields. On the basis of low-energy theories that we derive here,  it is remarkably simple to assess these nonlinear signatures for magnetic edge states. The effective theories are especially suitable in parameter regimes where other methods such as quantum Monte-Carlo simulations are exceedingly difficult due to exponentially small energy scales. 
The armchair ribbon setup discussed here provides a promisingly well-controlled (both experimentally and theoretically) environment for studying the principles behind edge magnetism in graphene-based nano-structures. 
\end{abstract}

\maketitle

Graphene~\cite{novoselov_2004}, a two-dimensional network of carbon atoms, has induced much excitement among physicists because of a multitude of unusual electronic properties~\cite{peres_rmp_graphene_2010}. Much of the literature about graphene has focussed on non-interacting electrons moving on a honeycomb lattice, though. One reason for this is that free electrons already show unorthodox effects such as Klein tunneling~\cite{young2009klein_tunneling,beenakker_klein_tunneling_rmp} and an anomalous quantum Hall sequence even at room temperature~\cite{novoselov2007room}. Another reason is that electron-electron interactions are suppressed close to the charge neutrality point because of the vanishing density of states, so that electronic correlation effects are often considered to be of minor importance in graphene.

At graphene edges the density of states may be peaked due to the presence of edge-localized states close to the Fermi level \cite{zigzag1}. Especially at extended zigzag edges this leads to a phenomenon called edge magnetism, where various theories predict ferromagnetic intra-edge and antiferromagnetic inter-edge correlations~\cite{zigzag2,zigzag3,zigzag4,zigzag5,zigzag6,zigzag7,zigzag8,zigzag9,zigzag9a,golor_2013}. Experimentally, these magnetic correlation effects prove to be elusive. Only recently, experimental indications of the importance of electron-electron interactions at edges of chiral graphene nanoribbons have been reported~\cite{crommie_2011}. The actual magnetic properties, however, remain unresolved experimentally as yet. The most severe issues hampering the experimental study of edge magnetism are (a) uncontrolled and rough edges~\cite{kunstmann_2011}, (b) hybridization with the substrate~\cite{yan_gnr_on_substrates_2012}, and (c) unclear experimental signatures of edge magnetism. All these issues are in fact related to the high complexity of the edges of most graphene nanoribbons: On the one hand the exact structural properties of the ribbon edges are not known. On the other hand, theory has not yet provided clear experimentally resolvable signatures beyond mean-field band structures.

Here,  we propose to study electronic correlation effects in a simpler geometry, namely in short armchair nanoribbons with zigzag ends (see Fig.~1). This geometry resolves at least issues (a) and (c). The basic principles of edge magnetism become strikingly clear in those ribbons, since, compared to edge magnetism in large zigzag or chiral ribbons, they offer three key advantages: (i) suitable high-quality armchair ribbons are already available \cite{muellen_termini,koch2012,swart}. (ii) As we will show, armchair ribbons are well under control theoretically as they allow for an essentially exact solutions without the need to resort to mean-field techniques. (iii) Correlation effects are accessible by means of magnetometry and spin-resolved scanning tunneling spectroscopy (STS).

\begin{figure}[!t]
\includegraphics[width=\linewidth]{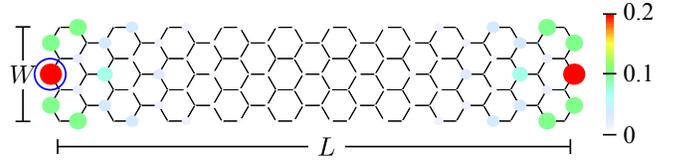}
\caption{(Color online) Lattice geometry of an armchair nanoribbon with $W=3$ hexagons in zigzag direction and $L=10$ hexagons in armchair direction. On top of the lattice the weight of the edge-localized low-energy states $|\psi_{\rm L}(i)|^2 + |\psi_{\rm R}(i)|^2$ is shown. Both the dot size and color scales with the weight. The blue circle indicates a typical site at which the spectral function is to be measured (see text).}
\label{fig_lattice_wfkt}
\end{figure}

{\it Model and geometry.} Our analysis is based on the $\pi$-band model of graphene. It is convenient for our purposes to separate the Hamiltonian $H = H_0 + H_U+ H'$ into a dominant part 
\begin{equation}
H_0 + H_U = -t\sum_{\langle i,j\rangle,\tau} (c^\dagger_{i\tau} c^{}_{j\tau}+ {\rm H.c.}) + U \sum_{i}c^\dagger_{i\uparrow}c^{}_{i\uparrow} c^\dagger_{i\downarrow} c^{}_{i\downarrow},
\end{equation}
where $\tau$ is a spin label, $t\approx 3\,$eV is the nearest-neighbor hopping amplitude on the honeycomb lattice and \mbox{$U\approx 6\,$eV}  the on-site Hubbard repulsion. $H'$ contains additional terms such as more distant neighbor hopping and the long-range part of the Coulomb repulsion. We will show that $H_0+H_U$ governs the physics, while $H'$ only renormalizes the effective parameters.
The lattice geometry that we consider is characterized by the number of hexagons in the zigzag direction, $W$, and the number of hexagons in the armchair direction, $L$ (see. Fig.~1). High quality nanoribbons with $W=3$ have recently been synthesized in a bottum-up approach in the laboratory~\cite{muellen_termini,koch2012,swart}. They have been shown to be terminated by single hydrogen atoms \cite{muellen_termini}, so that a $\pi$-band-only model is well justified.

{\it Effective low-energy theory.} Following Ref. \onlinecite{schmidt_eff_vs_qmc_2013} we derive a fermionic theory for the low-energy sector of a $W=3$ ribbon. The relevant degrees of freedom are selected on the basis of their localization properties. The eigenstates of $H_0$ are separated into bulk states $b^\dagger_{\mu\tau} = \sum_i\phi_\mu(i) c^\dagger_{i\tau}$ with $\sum_i |\phi_\mu(i)|^4 \sim 1/L$, and end states $e^\dagger_{\pm\tau} = \sum_i \psi_\pm(i) c_{i\tau}^\dagger$ with $\sum_i |\psi_\pm(i)|^4 \approx \rm const.$ for large $L$. The latter are symmetric and antisymmetric superpositions of exponentially localized states $d_{{\rm L/R},\tau}^\dagger$ at the left (L) and right (R) end of the ribbon. Typical end state wave functions are shown in Fig.~\ref{fig_lattice_wfkt}. We reconstruct the states $d^\dagger_{{\rm L/R},\tau}$ from the low-energy eigenstates $d_{\pm,\tau}^\dagger$ of $H_0$. The energy of the end states $\epsilon_+= -\epsilon_-$ are exponentially small in $L$ while we find that the bulk energies for Eq.~(1) satisfy $|\epsilon_\mu|\gtrsim 0.23 t$. Thus, the end states are energetically well separated from the bulk states. From the end state wave functions $\psi_{\rm L/R}(i)$ we construct a fermionic low-energy theory
\begin{multline}
H_{\rm f} = -t^* \sum_\tau (e^\dagger_{\rm L\tau}e^{}_{\rm R\tau} + {\rm H.c.}) -\!\!\!\!\!\!\sum_{\tau,\tau',s={\rm L,R}}\!\!\! e^\dagger_{s\tau} (\ve B \cdot \boldsymbol \sigma/2)^{}_{\tau\tau'} e^{}_{s\tau'} \\ + U^* \sum_{s={\rm L,R}} (e^\dagger_{s\uparrow} e^{}_{s\uparrow} -1/2)(e^\dagger_{s\downarrow} e^{}_{s\downarrow} -1/2), \label{eff_ferm}
\end{multline}
where $t^* = |\epsilon_\pm|$ describes an effective hopping of electrons from one end to the other, $U^* = U \sum_i |\psi_{\rm L}(i)|^4 = U \sum_i |\psi_{\rm R}(i)|^4 \approx 0.1\,U$ is an effective Hubbard repulsion for the end-localized electrons, $\boldsymbol\sigma$  the vector of Pauli matrices, and $\ve B$  an external magnetic field.

It is important to note, that $U^*$ is essentially independent of $L$ while $t^*$ becomes exponentially small for large $L$. By fitting the numerical data we find $t^* \approx e^{-L/1.86}\,1.29\,{\rm eV}$. Thus, for not too small $L$ a further reduction of $H_{\rm f}$ to a two-spin Heisenberg model
\begin{equation}
H_{\rm H} =  J_{\rm H} \ve S_{\rm L} \cdot \ve S_{\rm R} - \ve B\cdot (\ve S_{\rm L}+\ve S_{\rm R}),\bs J_{\rm H}=4(t^*)^2/U^*\label{heisenberg_model}
\end{equation}
is feasible. Here, $\ve S_{\rm L/R}$ are spin-$\frac12$ operators describing the spins of the localized electrons at the left/right end. This simple Heisenberg theory describes two antiferromagnetically coupled spins, localized at the ribbon ends, with a singlet-triplet (ST) splitting $J_{\rm H}>0$.

{\it Assessing the effective theory.} In order to scrutinize the effective low-energy theories (\ref{eff_ferm}) and (\ref{heisenberg_model}), we perform numerically exact projective auxiliary-field determinant quantum Monte-Carlo (QMC) simulations of the full lattice model $\tilde H = H_0+H_U$. Ground-state averages of arbitrary observables $\hat O$, such as the total energy or the Green's function, are calculated by $\langle\hat O \rangle=\langle\psi_{\rm T}|e^{-\theta \tilde H} \hat O e^{-\theta \tilde H}|\psi_{\rm T}\rangle / \langle\psi_{\rm T}|e^{-2\theta \tilde H}|\psi_{\rm T}\rangle$. The projection length $\theta=120/t$ is chosen sufficiently large as to ensure convergence. $|\psi_{\rm T}\rangle$ is taken as the ground state of the non-interacting system with a fixed number of spin-$\tau$ electrons $N_\tau$. We employ a third-order Trotter-Suzuki decomposition with a propagation step size of $\Delta\tau=0.01/t$. For further details on the QMC algorithm cf. Ref.~\onlinecite{assaad_det_qmc}.

\begin{figure}[!t]
\includegraphics[width=\linewidth]{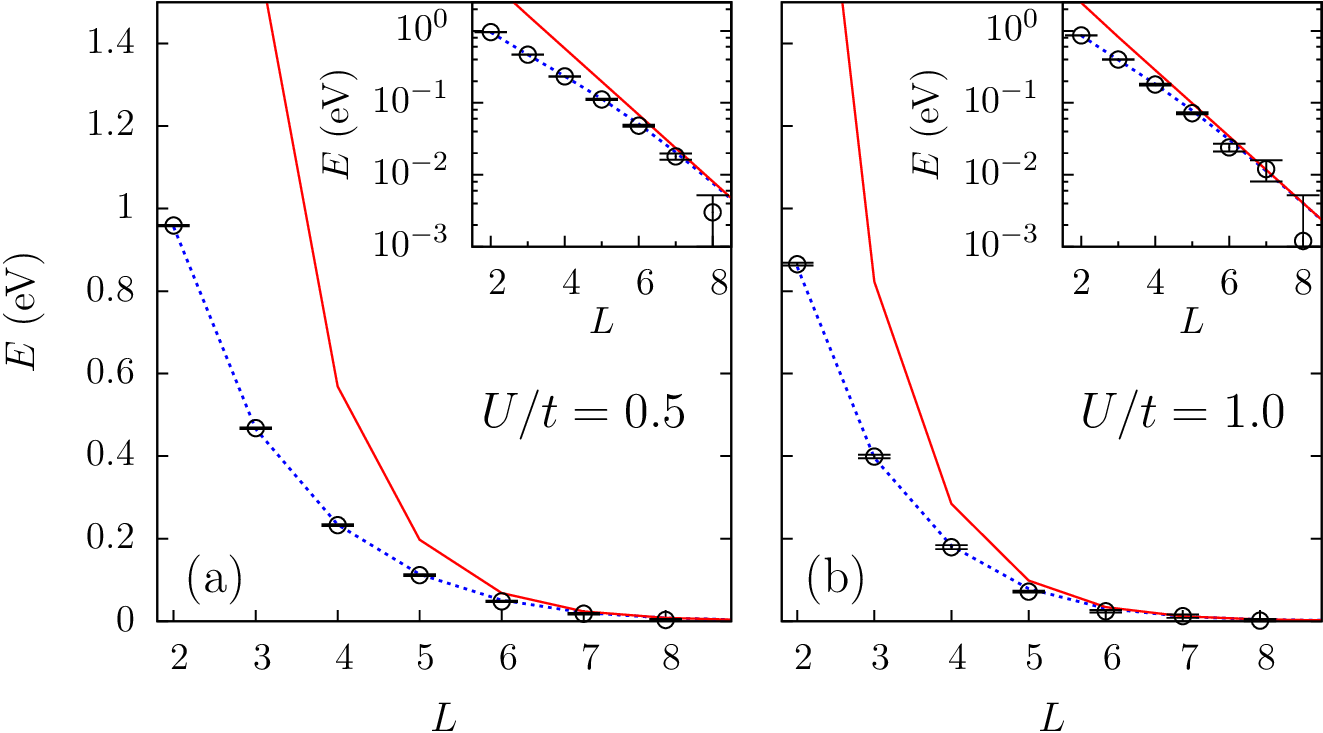}
\caption{(Color online) Singlet-triplet splitting $J$ as a function of ribbon length $L$. The blue dotted line is $J_{\rm f}$ calculated from $H_{\rm f}$ [Eq. (\ref{eff_ferm})]. The red solid line corresponds to $J_{\rm H}$ [Eq. (\ref{heisenberg_model})]. The black circles with error bars are the results of the QMC simulations $J_{\rm exact}$. (a) is with $U=0.5 t$ and (b) with $U=t$.}
\label{fig_eff_vs_qmc}
\end{figure}

Because of the $SU(2)$ symmetry, the single scale in the low-energy sector is the ST splitting, $J$, which we calculate in three different ways. (i) In the fermionic theory we find $J_{\rm f} = \sqrt{(U^*)^2/4+4(t^*)^2}-U^*/2$. (ii) In the Heisenberg theory $J_{\rm H}$ is given by Eq. (\ref{heisenberg_model}). (iii) Within QMC we calculate the difference $J_{\rm exact}$ of the total ground-state energies for $N_{\uparrow} = N_{\downarrow}+2$ (triplet sector) and $N_{\uparrow}=N_{\downarrow}$ (singlet sector). Figure \ref{fig_eff_vs_qmc} compares these three results for $U/t=0.5$ and $1.0$. While $J_{\rm f}$ agrees very well with the exact solution, $J_{\rm H}$ deviates significantly for very short ribbons, where $t^*/U^*$ is not yet small. Because of the exponentially small ST splitting, the QMC calculations were feasible only on relatively short ribbons $L\lesssim8$. In this regime, the fermionic theory agrees with the exact QMC results within error bounds. The effective theory, however, is not restricted to such small $L$. We conclude from Fig.~2, that for $L\gtrsim8$ the simple Heisenberg theory (\ref{heisenberg_model}) may be used to describe the spin physics.

{\it Detection via magnetometry.} The exponential dependence of $J$ on $L$ in combination with the discreteness of $L$ enables an experimental detection of the inter-end spin correlation by means of magnetization measurements. Note that $L$ must be even because of the synthesis process \cite{swart}. We assume an ensemble of $W=3$ ribbons of variable and homogeneously distributed sizes $L=8,10,\dots,20$ (corresponding to lengths between 4 and 8 nm), arranged randomly and sparsely on a 2D surface. Other length distributions can be accounted for easily. The surface of the largest ribbon is about $6\,$nm$^2$. Even if less than 1/10 of the substrate is covered by ribbons, $2\cdot 10^4$ ribbons per $\mu$m$^2$ are possible.

\begin{figure}[!t]
\includegraphics[width=\linewidth]{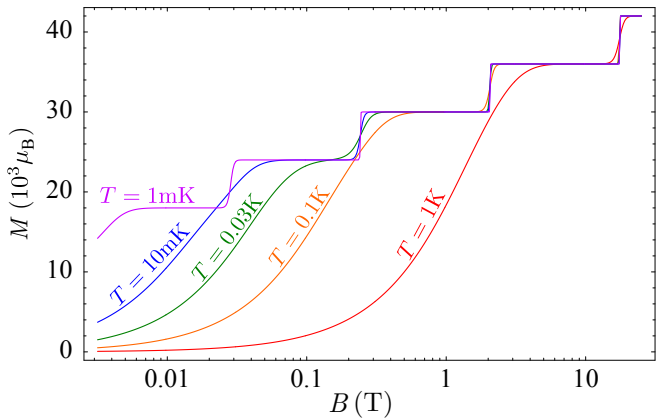}
\caption{(Color online) Total magnetic moment $M$ (in $10^3 \mu_{\rm B}$) of an ensemble of nanoribbons as a function of the magnetic field $B$ for different temperatures. The ensemble contains 3000 ribbons for each length $L=8,10,\dots,20$. Each individual ribbon contributes two Bohr magnetons $\mu_{\rm B}$ to the total moment if it is in the triplet state.}
\label{fig_ensemble_magnetization}
\end{figure}

In Fig. \ref{fig_ensemble_magnetization} we show the total magnetic moment of an ensemble of 3000 ribbons for each length $L=8,10,\dots,20$. At low temperatures one can see a clear non-linear response signature with steps of height $6000\,\mu_{\rm B}$. The spacing of the critical field strengths of the different steps corresponds roughly to one order of magnitude in the magnetic field. This non-linear response signature is observable with cutting-edge experimental magnetometry techniques \cite{hendrik_magnetometry}.

{\it Detection via spin-polarized STS.} A complementary experimental method for detecting the spin correlations is spin-polarized STS, which is capable of measuring the spin-resolved local spectral function
\begin{multline}
A_{i\tau}(\omega) = \frac1Z \sum_{m,n} |\langle m| c^\dagger_{i\tau} |n\rangle|^2 \\\times \left[ e^{-\beta E_n} + e^{-\beta E_m} \right] \delta_\eta\left(\omega - (E_m -E_n)\right).\label{spectralf}
\end{multline}
Here, $|n\rangle$ are manybody eigenstates of $H$ with energy $E_n$. Although the usual definition of the spectral function involves Dirac delta functions, we use the Gaussian $\delta_\eta(x)$ with finite width $\eta$. This accounts for a finite experimental energy resolution of the spectral function due to finite lifetimes or due to the temperature of the electrons and holes tunneling from the STM tip into the system. Note that this temperature $\eta$ is not necessarily equal to the ribbon temperature $\beta^{-1}$ in a non-equilibrium situation. For $i$ we choose a site at the ribbon edge, where one of the edge state wave functions is large (see Fig. \ref{fig_lattice_wfkt}). Here we opt for the site $i$ on the left.

\begin{figure}[!t]
\includegraphics[width=\linewidth]{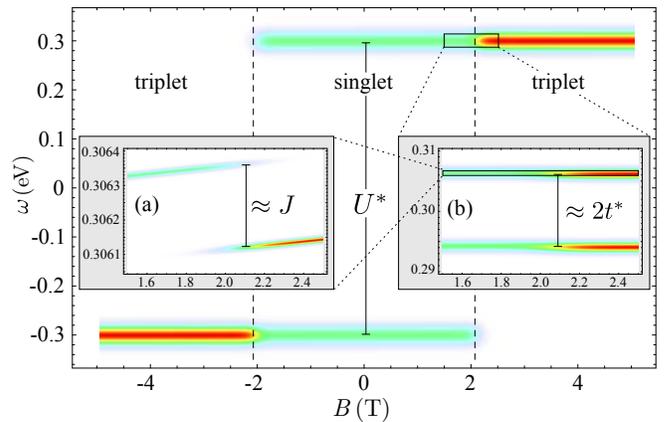}
\caption{(Color online) Spectral function $A_{{\rm L}\downarrow}(\omega)$ for $t^*=0.006$ eV and $U^* = 0.6$ eV, corresponding to a ribbon with $W=3$ and $L=10$. The critical field $B_c = \pm J/2 \approx 2.1\,$T is indicated by dashed lines. In the main plot the Gaussian smearing $\eta=200$ Kelvin (see text). The insets show the fine structure of the spectral peaks with narrower Gaussian smearing of $\eta = 10$ Kelvin (b) and $\eta=0.1$ Kelvin (a).}
\label{fig_spectral_function}
\end{figure}

The lattice electron operator $c^{}_{i\tau}$ may be expressed in edge and bulk states. But since we are only interested in small energies $\omega$ we may drop the bulk states and obtain $A_{i\tau} (\omega) = |\psi_{\rm L}(i)|^2 A_{{\rm L}\tau}$, where $A_{{\rm L}\tau}(\omega)$ is of the same form as Eq. (\ref{spectralf}), with the $c^{}_{i\tau}$ operator replaced by $e^{}_{{\rm L}\tau}$. We have assumed that $i$ is on the left end of the ribbon. We calculate the spectral function within the fermionic effective theory [Eq. (\ref{eff_ferm})] by exact diagonalization. The resulting exact spectral function $A_{\rm L\downarrow}(\omega)$ in dependence on the external magnetic field $B$ is shown in Fig. \ref{fig_spectral_function}. The large energy features provide a clear distinction between a singlet phase for $|B|<B_c$ and two triplet phases for $|B|>B_c$. For a better understanding of these large energy features we evaluate $A_{{\rm L}\tau}(\omega)$ approximately for $U^* \gg B,T,t^*$ and find
\begin{equation}
A_{{\rm L}\tau}(\omega) = \frac12\sum_{\tau'}\delta_\eta\!\left(\omega + \tau' \frac{U^*}2\right) \frac{1+e^{\beta J}+2e^{2\tau\tau'\beta B}}{1+ e^{\beta J}+2\cosh(2\beta B) }.\label{spectrum_approximation}
\end{equation}
This approximate formula only holds for $\omega$ resolutions $\eta$ coarser than the energy scales $J,t^*$. It shows that for $|B|<B_c=J/2$ it is possible to add a spin-down electron with energy $U^*/2$ and to remove a spin-down electron with energy $-U^*/2$, because the singlet phase is a coherent superposition of a state with an up-spin and a state with a down-spin on the left side. As  $B>B_c=J/2$ the triplet state $e^\dagger_{{\rm L}\uparrow} e^\dagger_{{\rm R}\uparrow}|0\rangle$ becomes the ground state, where it is only possible to add an down-spin electron but not to remove one. This is reflected by all the spectral weight being at positive $\omega$. We also calculated the exact spectral function by QMC (not shown in Fig. \ref{fig_spectral_function}) and found that, within the statistical bounds achievable in QMC, it agrees with Eq.~(\ref{spectrum_approximation}) in the low-energy regime $|\omega| \lesssim U^*/2$.

In addition to the large energy features there is a subtle fine structure, shown in the insets of Fig. \ref{fig_spectral_function}. Interestingly, the fine structure reflects all energy scales appearing in the effective theories, i.e., the inter-end hopping $t^*$ and, on an even finer level, the antiferromagnetic inter-edge coupling $J = 4(t^*)^2/U^*$. For different geometries, resulting in different effective model parameters, the energy scales of the fine structures scale accordingly.

{\it Stability with respect to $H'$.} The fermionic low-energy theory [Eq.~(\ref{eff_ferm})] is derived by first order perturbation theory from $H_0 +H_U$, with the inverse bulk gap being the small parameter. As shown in Fig. \ref{fig_eff_vs_qmc}, this is a remarkably good approximation. Moreover, due to overall SU(2) invariance, the low-energy spin physics must be governed by one single energy scale, namely the singlet-triplet splitting $J$. Thus it is reasonable to account for the additional terms $H'$ in the effective Heisenberg model also in first order perturbation theory and calculate the corrections to $J$. We now discuss a variety of possible perturbation terms in $H'$. We find that the resulting corrections to $J$ are smaller than the uncertainty for the literature parameters $t$ and $U$. 

In the special geometry discussed here, next-nearest neighbor hopping only gives rise to a shift in the chemical potential and may therefore be ignored. Third neighbor hopping and all other hoppings that couple the A and B sublattice give contributions to $t^*$ which are at least one order of magnitude smaller than the contribution of the nearest-neighbor hopping, since these more distant hoppings are exponentially suppressed as a function of distance. An electric field $E$ along the ribbon gives rise to a correction $\Delta J = J (E L/2U^*)^2$.

Two additional terms arise from the long-range part of the Coulomb interaction. One one hand there is the inter-edge term $V_{\rm LR} n_{\rm L} n_{\rm R}$ with $n_{s} = \sum_\tau e^\dagger_{s\tau} e^{}_{s\tau}-1$, $V_{\rm LR} \approx e^2\exp(-L/L_{\rm sc})/\kappa l$ with $\kappa$ the dielectric constant and $L_{\rm sc}$ the charge screening length, both of which depend heavily on the substrate. On the other hand there is an intra-edge term $V_0  \sum_{s} (e^\dagger_{s\uparrow} e^{}_{s\uparrow} -1/2)(e^\dagger_{s\downarrow} e^{}_{s\downarrow} -1/2)$ with $V_0 \approx e^2 \sum_{i<j} |\psi_{\rm L}(i)|^2 |\psi_{\rm L}(j)|^2 \exp(-{|\ve r_i-\ve r_j|}/L_{\rm sc})/ {\kappa |\ve r_i-\ve r_j|}$, which has the same form as the Hubbard term in Eq.~(\ref{eff_ferm}). Both terms contribute to the energy of the excited states with more or less than one electron at one end and therewith to $U^*$. Given the uncertainties for $U$ in the literature (see, e.g., Ref. \onlinecite{schuler2013hubbard} for an interesting discussion),  the actual value of the renormalized $U^*$ with all corrections due to the environment is not accessible theoretically and needs to be determined experimentally. One way to do so is via a measurement of the spectral function (Fig. \ref{fig_spectral_function}) in tunneling experiments. Crucially, however, the exponential length dependence of $J$ is not spoiled by any of the perturbations to $H_0+H_U$ discussed above.

{\it Conclusion.} Motivated by the recent chemical synthesis of high-quality armchair ribbons with perfect zigzag ends \cite{muellen_termini,koch2012,swart}, we have studied the magnetic correlations arising in these ribbons due to strong electronic interactions. We have identified spin-$\frac12$ degrees of freedom at each end of the ribbon. This spin subsystem may be described by an extremely simple Heisenberg model, which we have derived directly from a lattice model and benchmarked against numerically exact QMC simulations. The two end spins are coupled antiferromagnetically and the corresponding singlet-triplet splitting decays exponentially with the ribbon length. This enables direct experimental access to the low-energy spin physics, for which we have proposed two complementary experiments. The  setup proposed here thus allows to investigate the basic principles of graphene edge magnetism. A thorough understanding of this well-controlled scenario will facilitate the experimental investigation as well as the theoretical interpretation of the still-elusive edge magnetism in larger zigzag/chiral nanoribbons.

We want to thank H. Bluhm and M. Morgenstern for their valuable remarks on the feasibility of the experiments proposed here. Furthermore, we thank J. van der Lit and I. Swart for insightful discussions and for sharing unpublished results with us. Financial support by the DFG under Grant WE 3649/3-1 and FOR 1807 is gratefully acknowledged, as well as the allocation of CPU time within JARA-HPC and from JSC J\"ulich.

\bibliography{refs}

\end{document}